\documentclass[a4paper,10pt]{article}
\usepackage[margin=2cm]{geometry}
\usepackage[english]{babel}
\usepackage[latin1]{inputenc} 
\usepackage[pdftex]{graphicx} 
\usepackage[round]{natbib}

\newcommand{\dfrac}[2]{\frac{\displaystyle#1}{\displaystyle#2}}

\title{Dynamics of solar wind protons reflected by the Moon}
\author{M.\ Holmstr{\"o}m\thanks{Swedish Institute of Space Physics, PO~Box~812, SE-98128~Kiruna, Sweden. (matsh@irf.se)}, 
M.\ Wieser$^*$, 
S.\ Barabash, 
Y.\ Futaana, 
and A.\ Bhardwaj\thanks{Space Physics Laboratory, Vikram Sarabhai Space Center, Trivandrum, India}}

\begin{document} 
\maketitle

\begin{abstract}
Solar system bodies that lack a significant atmosphere and significant 
internal magnetic fields, such as the Moon and asteroids, have been 
considered as passive absorbers of the solar wind. 
However, ion observations near the Moon by the SELENE 
spacecraft show that a fraction of the impacting 
solar wind protons are reflected by the surface of the Moon. 
Using new observations of the velocity spectrum of these 
reflected protons by the SARA experiment 
on-board the Chandrayaan-1 spacecraft at the Moon, 
we show by modeling that the reflection of solar wind protons will 
affect the global plasma environment. 
These global perturbations of the ion fluxes and the magnetic fields will depend on 
microscopic properties of the object's reflecting surface. 
This solar wind reflection process could explain past ion 
observations at the Moon, and the process should 
occur universally at all atmosphereless non-magnetized objects.  
\end{abstract}

\section{Introduction}
Traditionally, bodies that lack a significant atmosphere and internal 
magnetic fields, such as the Moon and asteroids, have been considered 
passive absorbers of the solar wind~\citep{Cravens}. 
The solar wind ions and electrons directly impact the surface of these bodies 
due to the lack of atmosphere, and the interplanetary magnetic field passes 
through the obstacle relatively undisturbed because the bodies are 
assumed to be non-conductive. 
Since the solar wind is absorbed by the body, 
a wake is created behind the object. 
This wake is gradually filled by solar wind plasma downstream of the body, 
through thermal expansion and the resulting ambipolar electric field, 
along the magnetic field lines~\citep{Farrell98}, 
This picture of the interaction between the Moon (and asteroids) 
and the solar wind, is based on in-situ observations of ions, electrons, 
and magnetic fields 
by many missions~\citep{shu74,Ogilvie96,Halekas05,Nishino09a}. 

However, there have been observations that do not easily 
fit into this picture of atmosphereless bodies as the 
passive absorbers of the solar wind. 

On the Moon, the Apollo~12 and~14 Suprathermal Ion Detector (SIDE) 
observed energetic ion fluxes at the nightside surface~\citep{fre72}. 
Also, Nozomi observed non-thermal ions at large distances from, 
and upstream of, the Moon~\citep{Fut03}.  Such ions are not easily 
explained in the traditional picture of the Moon--solar wind interaction. 

Recent observations at the Moon by the SELENE (Selenological and 
Engineering Explorer) mission~\citep{KaguyaGRL08} 
and by the Chandrayaan-1 mission~\citep{gos09} 
might provide a clue to many of these unexplained observations. 
Ion detectors on-board SELENE observed 
that some of the solar wind protons (around 0.1\%) are 
reflected by the Lunar surface, which was unexpected, 
since it has been assumed that all solar wind protons are neutralized 
on impact with the Lunar surface~\citep{cri02}. 

\begin{figure}
\begin{center}
  \includegraphics[width=0.7\columnwidth]{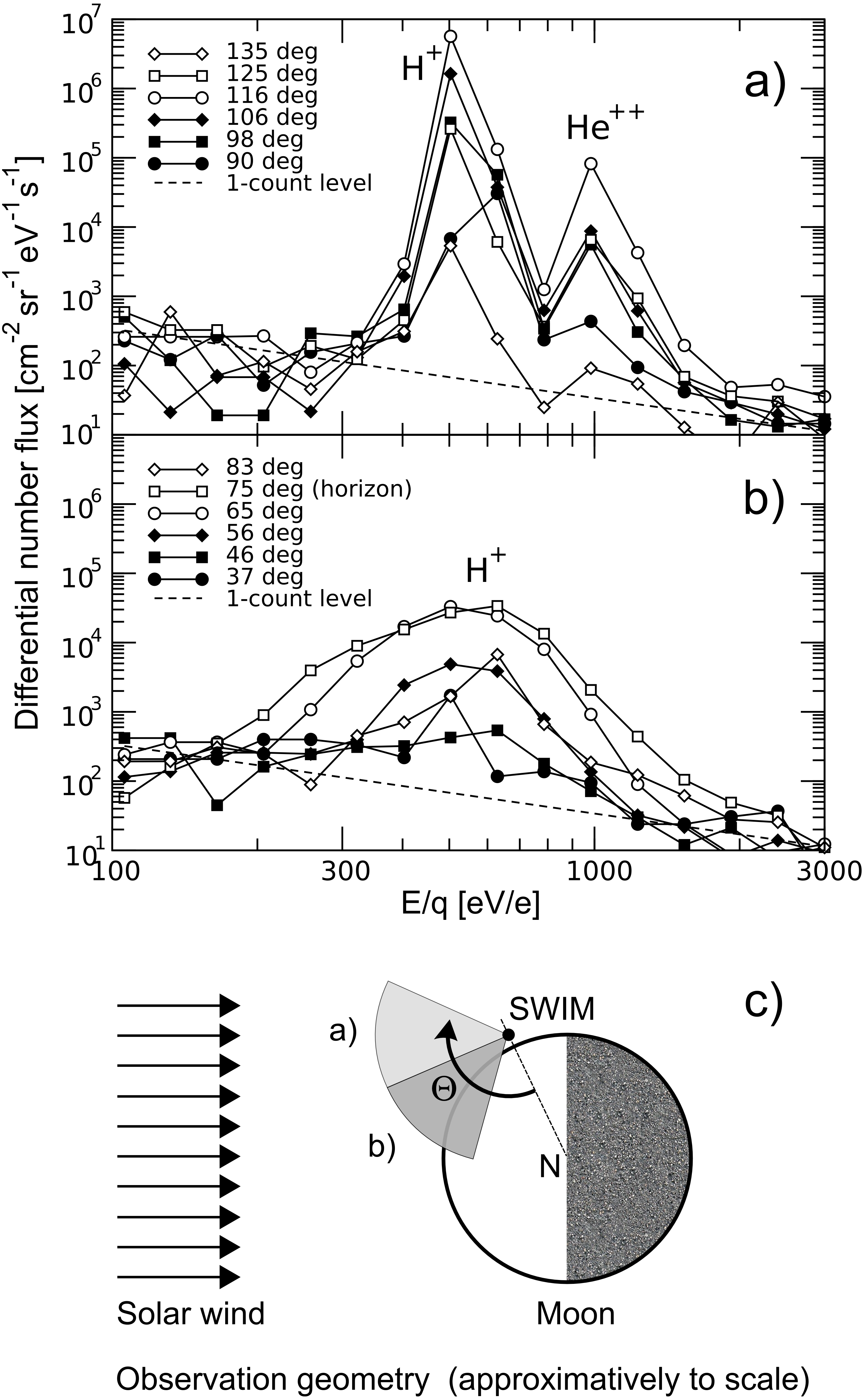}
\end{center}
\caption{Ion observations on February 19, 2009, by the SWIM sensor, 
of the SARA instrument, on Chandrayaan-1.  
In (a) and (b) we show ion energy spectra from different directions, 
and in (c) the observation geometry is illustrated.  
The view directions are approximately in the equatorial plane, 
and they are labeled by the angle, $\Theta$, to the nadir direction.  
The spectra in (a) show the undisturbed
solar wind from 106 and 116 deg, while the spectra in (b) show the protons 
reflected from the surface. 
The spectra are averaged over 30~minutes, and 
are from twelve of the 16~sectors (the remaining four 
sectors show only background values).
The plane of the illustration in (c) is approximately the 
equatorial plane.   Chandrayaan-1 is in a circular polar orbit 
at 100~km altitude~\citep{gos09}. 
The SWIM sensor has a total field of view of 
$7.5^{\circ}$ FWHM $\times 180^{\circ}$ divided into 16 sectors, and a 
time resolution of 8 seconds~\citep{swim}. 
The energy resolution is $dE/E=0.07-0.08$, depending on viewing direction. 
} \label{fig:ch1}
\end{figure}

Also, the ion detector (SWIM) of the SARA experiment on-board the 
Chandrayaan-1 mission~\citep{bha05,swim,bar09,wie09} 
observes these reflected solar wind protons, and in 
Fig.~\ref{fig:ch1} we show the energy spectrum of one such observation. 
The sunward sectors of the detector see an undisturbed solar wind, 
while the surface looking sectors see reflected protons with a 
broader energy distribution. 
These observations show that the simplified picture of atmosphereless 
bodies as passive absorbers of the solar wind is incomplete. 

The reflection of solar wind ions on 
solar system bodies that lack a significant atmosphere 
affects the solar wind interaction, where 
the microphysical properties of the reflecting surface 
will perturb the global ion and magnetic field environment 
near the object.  
We illustrate this effect on the solar wind interaction 
by modeling results for ions near the Moon. 
In particular, we show that this model can explain the SARA/Chandrayaan-1 ion 
observations (Fig.~\ref{fig:ch1}) and the observations of 
non-thermal ions by the Nozomi mission~\citep{Fut03}.

\section{Model}
We present a model of the surface reflection of solar wind protons at the 
Moon and its effects on the global ion distribution in the vicinity of 
the Moon.  Reflection (or back scattering) of solar wind protons on the 
surface of a solar system body is a process where 
some of the solar wind protons that impact the surface 
will recoil on atoms in the top atomic layers of the regolith, 
with only a slight reduction in velocity. 

\begin{figure}
\begin{center}
  \includegraphics[width=0.55\columnwidth]{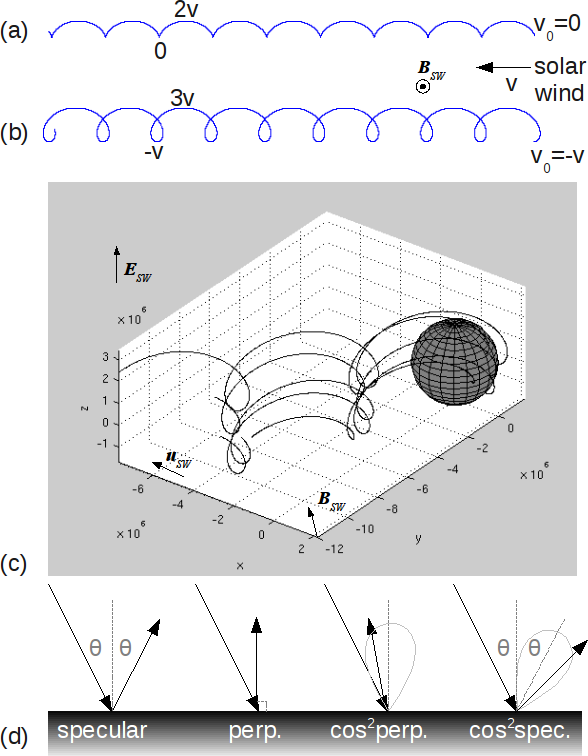}
\end{center}
\caption{(a) The cycloid trajectory of a pick-up ion where the initial 
velocity is zero, and a solar wind of velocity $v$ is inflowing from the 
right.  The solar wind magnetic field, $\mathbf{B}_{sw}$, is directed out of 
the page.  The solar wind convectional electric field is 
$\mathbf{E}_{sw} = \mathbf{v}\times\mathbf{B}_{sw}$. 
The ion velocity vary between $0$ and $2v$ along the trajectory. 
(b) The trajectory of an ion injected into the solar wind with velocity 
$-v$.  This is equivalent to a pick-up ion with zero velocity in a solar wind 
of velocity $2v$, as seen by an observer moving with velocity $-v$. 
In the frame of the illustration, where the plasma flow to the left with 
velocity $v$, the ion velocity will vary between $-v$ and $3v$ 
along the trajectory. 
(c) Sample trajectories of protons launched 
perpendicular to the lunar surface.  
(d) Illustration of some different reflection models, i.e.\ different ways 
of selecting the direction of the velocity vector of a proton that has 
reflected on the planetary surface.  
} \label{fig:ill}
% (a,b) /home/matsh/shared/projects/phobos/traj/cyclo.m
% (c)   /home/matsh/shared/projects/phobos/traj/trajs.m
\end{figure}

We assume that there are three basic parameters that characterize 
the reflection process, 
(1) the fraction of the precipitating protons that are reflected, $f_r$, 
i.e.\ the probability that a proton is reflected.  We assume here that it 
is a constant, independent of impact velocity. 
(2) the speed of a reflected protons, as a fraction of the 
impact speed, $f_v$, and 
(3) the directional (angular) distribution of the reflected protons. 
The first published observation of reflected protons was by 
SELENE at the Moon~\citep{KaguyaGRL08}. 
Their estimate is that the reflected fraction $f_r=0.001-0.01$
(this is consistent with estimates from SWIM observations), 
and that the velocity magnitude of the reflected protons is 80\% of the 
solar wind velocity magnitude, $f_v=0.8$. 
Regarding the angular distribution of the reflected protons, 
\citet{KaguyaGRL08} find that it is much broader than that of the 
solar wind ions, thus the observed ions are not 
specularly reflected but rather scattered at the lunar surface. 
Since the exact angular distribution is unknown, we have used 
four different reflection models for 
determining the velocity direction of the reflected ion. 
\begin{itemize}
\item Specular.  For specular reflections, the protons' velocity vector 
      before and after the reflection are in the same plane, and the angle 
      to the surface normal is the same. 
      So, an ion inside the spherical obstacle at the position 
      $\mathbf{r}$, with velocity $\mathbf{v}$, is reflected by updating 
      the velocity to $\mathbf{v'} = \mathbf{v} - 2\left(\mathbf{v}\cdot 
      \mathbf{\hat{r}}\right)\mathbf{\hat{r}}$, where 
      $\mathbf{\hat{r}}=\mathbf{r}/|\mathbf{r}|$. 
\item Perpendicular, $\mathbf{v'}$ is perpendicular to the surface of the 
      spherical obstacle (parallel to the surface normal). 
\item cos$^2$-perpendicular.  The angle between $\mathbf{v'}$ and the 
      surface normal, $\theta$, is randomly drawn from a $\cos^2\theta$ 
      probability distribution. 
\item cos$^2$-specular. The angle between $\mathbf{v'}$ and the 
      direction of specular reflection (see above), 
      $\theta$, is randomly drawn from a $\cos^2\theta$ 
      probability distribution. 
\end{itemize}
These different reflection models are illustrated in Fig.~\ref{fig:ill}d. 
If not noted otherwise, we have used the cos$^2$-specular reflection model 
since that gave the best fit to the observations. 

When a solar wind proton has reflected, it will travel in a 
cycloid motion, gyrating around the magnetic field lines. 
The motion of the ion with charge $q_i$ and velocity 
$\mathbf{v}_i=\mathbf{v}_i(t)$ is governed by the Lorentz force, 
\[
q_i \left( \mathbf{E}+\mathbf{v}_i\times\mathbf{B} \right) 
= q_i ( \mathbf{v}_i-\mathbf{v}_{sw} ) \times\mathbf{B}_{sw}
\]
since the solar wind convective electric field is 
$\mathbf{E}_{sw}=-\mathbf{v}_{sw}\times\mathbf{B}_{sw}$. 
Here $\mathbf{v}_{sw}$ and $\mathbf{B}_{sw}$ is the solar wind velocity, 
and the IMF, respectively. 
We have assumed constant solar wind conditions. 

For an ion with zero initial velocity, this lead to the classical 
trajectory of a pick-up ion as illustrated in Fig.~\ref{fig:ill}a.  
Since the reflected protons have non-zero velocity 
in a coordinate system where the reflecting body is stationary, 
the trajectory will be different, as shown in Fig.~\ref{fig:ill}b.
In addition to the cycloid motion perpendicular to the IMF, 
the ion will also drift with a constant velocity along the IMF, 
if the initial velocity of the ion had a non-zero velocity 
component along the IMF.  
The trajectories of five ions, 
launched perpendicular to the Lunar surface at (lng,lat) = (0,0) and 
(+-30,+-30) degrees with velocities of 175~km/s 
in a 350~km/s solar wind are shown in Fig.~\ref{fig:ill}c as an 
illustration of this combined cycloid and drift motion. 
The solar wind magnetic field 
has a magnitude of 3~nT and is directed along $(-1,1,0)$. 
Since the launched protons have non-zero velocity components along the 
magnetic field they will travel along the field line in addition to the 
ideal cycloid motion shown in Fig.~\ref{fig:ill}b. 
The coordinate system has the $x$-axis from the center of the Moon 
toward the Sun, the $y$-axis in the ecliptic plane, and the $z$-axis 
in the northern ecliptic hemisphere. 

Previous investigations of ion trajectories near the Moon have 
considered ions produced by photoionization of exospheric 
neutrals~\citep{Manka70,Manka73}, not reflected ions. 
Neither has previous global models of the Moon's interaction with the 
solar wind included reflected solar wind protons 
\citep{kimura08,tra05,kal05,har03}. 
However, in a more general setting, \citet{Shimazu99} 
saw acceleration of reflected ions in hybrid simulations of plasma flow 
around a generic obstacle. 

\begin{figure}
\begin{center}
  \includegraphics[width=0.95\columnwidth]{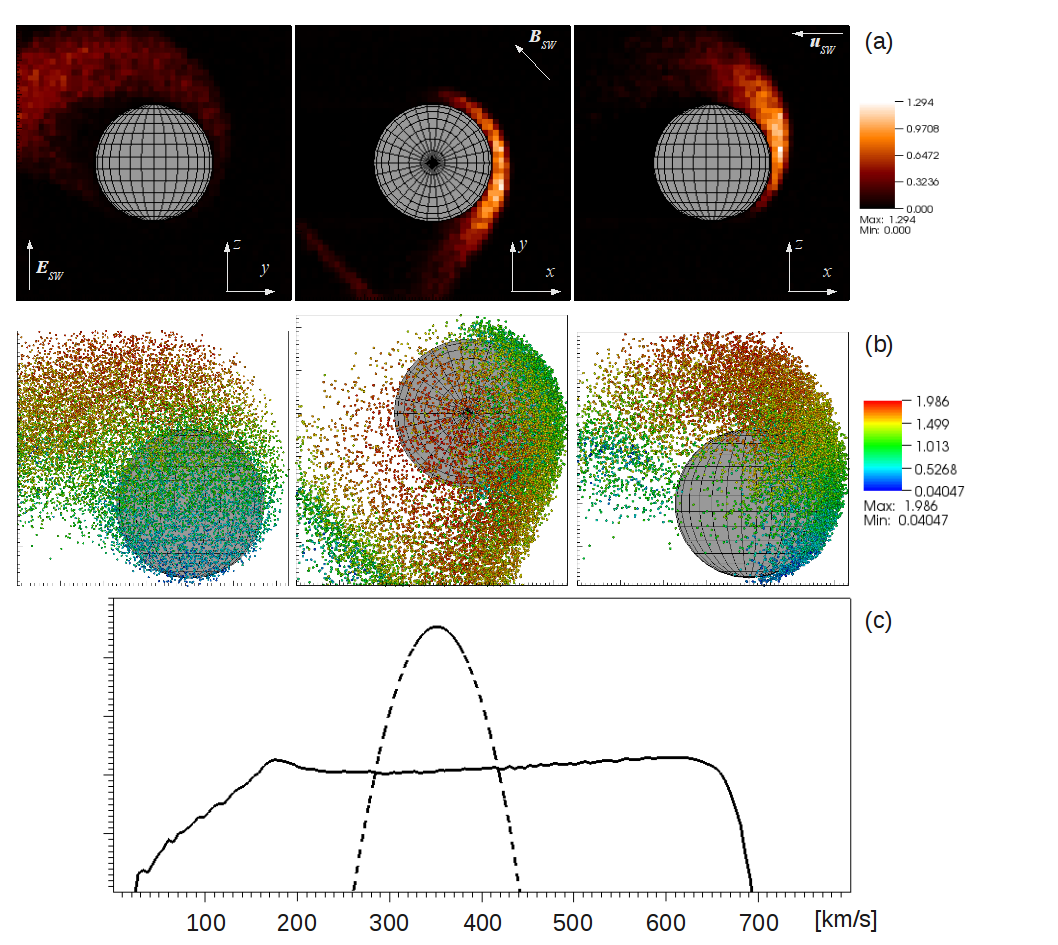}
  % /opt2/run/hpc2n/mb00 (mb10 has better resolution, but no particles)
  % (a) mb00
\end{center}
\caption{
Results for a test particle simulation run. 
(a) Magnitude of the $yz$-component of the proton number flux 
(number density times average velocity) 
around the Moon, relative to the magnitude of the solar wind proton 
number flux.  This is for the case of 100\% reflection ($f_r=1$) of the 
precipitating protons, but can be scaled for any other value of $f=1$. 
Shown are cuts through the planes $x=0$ (left), $z=0$ (middle), 
and $y=0$ (right). 
(b) The reflected proton meta particles, colored by velocity magnitude 
(relative to the solar wind velocity).  
%Same views as in (a). 
(c) The velocity spectrum of all protons in the simulation domain. 
The $x$-axis unit is~km/s, and the $y$-axis scale is logarithmic in 
arbitrary units. 
The solid line is the reflected protons, 
and the dashed line is the solar wind protons. 
        } \label{fig:flux}
% mb00
\end{figure}

For the modeling of reflected protons at the Moon we have used a 
test particle approach, where the trajectory of each proton is 
computed by integrating the Lorenz' force for constant solar 
wind conditions, i.e.\ constant solar wind conditions throughout 
the simulations domain. 
In what follows, the coordinate system used is centered at the Moon and 
has its $x$-axis toward the Sun, 
a $z$-axis perpendicular to the ecliptic plane, in the northern hemisphere, 
and a $y$-axis that completes the right handed system.  
% Parameters from /opt2/run/hpc2n/mb00/flash.par
The outer boundary of the simulation domain is a box centered at the Moon 
with sides of length 8000~km. 
The inner boundary is a sphere of radius 1730~km. 
At the start of the simulation the domain is empty of particles. 
Before each time step we fill the $x$-axis shadow cells (cells just 
outside the simulation domain) with proton meta-particles with 
weight $N_m = 7.5\cdot 10^{20}$ (number of real protons per meta-particle).  
The proton meta-particles are drawn from a Maxwellian distribution 
with a specified temperature and bulk velocity. 
After each time step the shadow cells 
and the obstacle region are emptied of protons.  
A fraction, $f_r$, of the protons found inside the obstacle are selected to be 
reflected, randomly, 
with a velocity magnitude that is reduced by a fraction $f_v$. 
The boundary conditions in the $y$- and $z$-directions are periodic. 
We have $N_I$ (meta-)ions at positions $\mathbf{r}_i(t)$~[m] with velocities 
$\mathbf{v}_i(t)$~[m/s], mass $m_i$~[kg] and charge $q_i$~[C], 
$i=1,\ldots,N_I$. 
Using a Leap-Frog integrator, 
the trajectories of the ions are computed from the Lorentz force, 
\[
  \dfrac{d\mathbf{r}_i}{dt} = \mathbf{v}_i, \quad
  \dfrac{d\mathbf{v}_i}{dt} = \dfrac{q_i}{m_i} \left( 
    \mathbf{E}+\mathbf{v}_i\times\mathbf{B} \right), \quad
    i=1,\ldots,N_I
\]
where $\mathbf{E}=\mathbf{E}(\mathbf{r},t)$ is the electric field, 
and $\mathbf{B}=\mathbf{B}(\mathbf{r},t)$ 
is the magnetic field, and time is stepped forward by 0.05~s
In Fig.~\ref{fig:flux} we show results for a model run 
with a solar wind with a velocity of
350~km/s, a temperature of 48000~K, and a magnetic field that is 
$(-2.12,2.12,0)$~nT, at time 20~s (when the solution has reached a 
steady state). 
The velocity of the reflected protons is reduced by $f_v=0.5$. 
Note that this is for the case of complete reflection of all 
precipitating protons ($f_r=1$), but this can be scaled for other 
reflection fractions, e.g., by 1/100 for a 1\% reflection. 

\begin{figure}
  % p3d35n and mb00
  % Should I use mb10d instead?
\begin{center}
  \includegraphics[width=1.0\columnwidth]{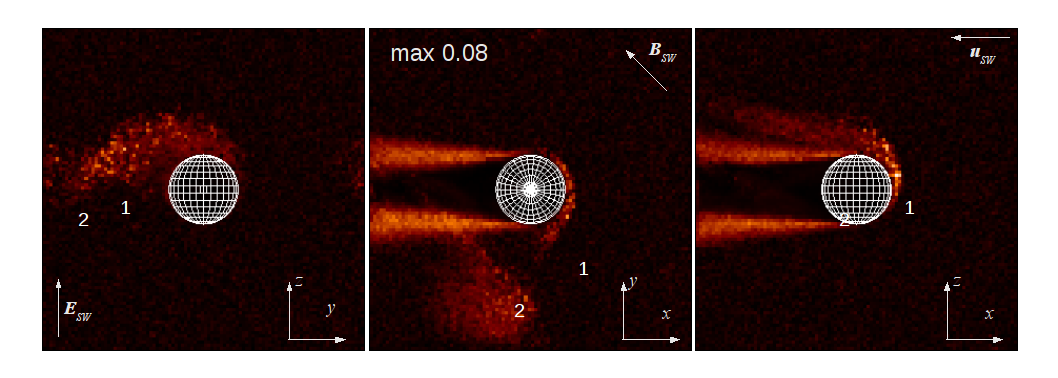}
\end{center}
\caption{Solar wind proton reflection using a hybrid model. 
Shown is the magnitude of the $yz$-component of the proton number flux, 
as in Fig.~\ref{fig:flux}a.  The test particle run is the same as in 
Fig.~\ref{fig:flux}.  The additional parameters for the hybrid run is 
$\mathbf{n}_{sw}=5$~cm$^{-3}$, $\mathbf{T}_{e}=10000$~K, $\gamma=5/3$, 
$\Delta t=0.08$~s, five subcycles per time step, $f_r=0.05$, approximately 
8 million meta-protons, 
the simulation domain is twice as large (a cube with sides 16000~km), 
and the grid has 100 cells along each dimension (one million cells in total). 
Also shown, by numbers, is the position of the two Nozomi observations that 
are presented in Fig.~\ref{fig:nozomi}. 
        } \label{fig:hybrid}
\end{figure}

In reality the solar wind flow is affected by the presence of the 
Moon and the reflected protons.   To correctly model this a 
self-consistent model is needed, at a considerable computational cost. 
Also, since this is a first attempt to model the effects of reflected 
protons, and the details of the reflection process is not known, 
a fully self consistent model of the process would not 
add much to the investigation. 
To justify the use of the test particle approach we did a 
self-consistent hybrid model~\citep{Enumath09} run using the same parameter 
values as used in Fig.~\ref{fig:flux}.  
The result is shown in Fig.~\ref{fig:hybrid}.  
We see that the morphology of the reflected ion fluxes are similar, 
although the maximum flux is smaller in the test particle case 
(0.065 times the solar wind flux, i.e. 1.3 times the reflection of 
0.05 used in the hybrid model) compared to the hybrid model (0.08), probably 
due to statistical fluctuations in the hybrid simulation. 
In the hybrid simulation we also see the wake refill behind the 
Moon, which is not present in the test particle simulation, but that 
is acceptable, since we are mostly interested in the dayside 
dynamics of the reflected solar wind protons.

\section{Model Results and Comparison with Observations}
We now investigate in more detail the general morphology of the 
reflected ions. 
The proton flux, in directions perpendicular to the solar wind flow direction, 
around the Moon is shown in Fig.~\ref{fig:flux}a. 
The global effects of the reflected protons are clearly visible. 
The reflected solar wind protons create a plume of ions that initially are 
accelerated along the solar wind electric field direction, then drift along 
the IMF direction, and gyrate perpendicular to the IMF. 
The density of reflected ions is highest near the sub solar point, 
then decrease tailward from dispersion due to different initial 
velocities. At the cusps of the cycloid motion there is however a 
density increase, as seen in the lower left corner of the middle plot 
of Fig.~\ref{fig:flux}a. The maximum density is 1.3 times the 
solar wind density when all protons are reflected.  So if 1\% of 
the solar wind protons are reflected the maximum density would be 
0.013 times the solar wind density. 

\begin{figure}
\begin{center}
  \includegraphics[width=0.8\columnwidth]{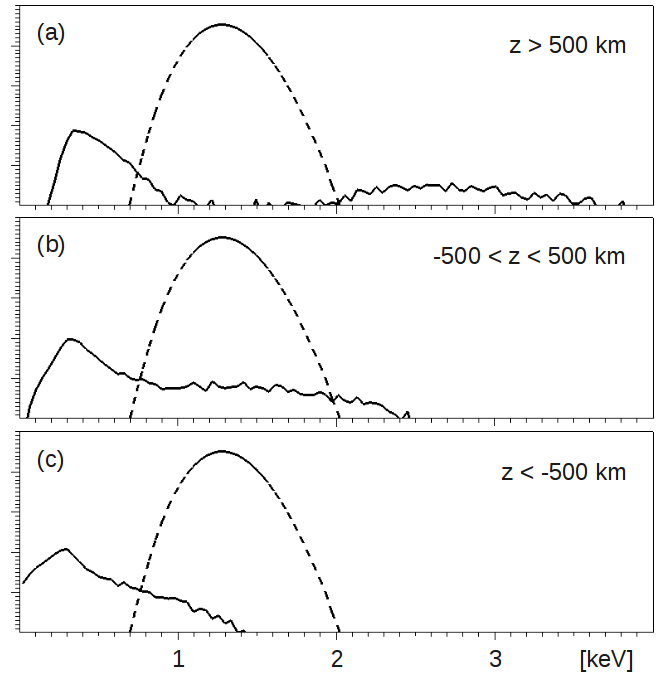}
\end{center}
\caption{
Energy spectra of reflected protons (solid lines) 
  and solar wind protons (dashed lines). 
Corresponding to the spectrum in Fig.~3c, but instead of 
including all protons in the domain, each spectrum is for selected 
regions of the simulation domain. 
The included protons are those on the dayside ($x>0$) at altitudes less 
than 200~km with $z$ coordinates in the ranges 
(a) $z > 500$~km, 
(b) -500 $< z < 500$~km, and 
(c) $z < -500$~km. 
The solid line is the reflected protons, 
and the dashed line is the solar wind protons. 
The $y$-axes are logarithmic in arbitrary units. 
} \label{fig:spec}
\end{figure}

In Fig.~\ref{fig:flux}b is shown the meta-particles that correspond 
to the reflected ions.  The cycloid motion, and acceleration along 
$\mathbf{E}_{sw}$ is again clearly visible. 
The velocity spectrum of all protons in the simulation domain in 
Fig.~\ref{fig:flux}c, shows the solar wind population, and the much 
broader distribution of reflected ions. 
This broadening of the spectrum is consistent with the 
broadening of the observed spectrum for look direction toward the 
lunar surface, as shown in Fig.~\ref{fig:ch1}b. 
The similarity between the observed and modeled spectrum is 
even better if we only include protons in the model that are 
close to the position of Chandrayaan-1 at the observation time.  
In Fig.~\ref{fig:spec} we show the spectra of reflected protons in 
three latitude bands, with Fig.~\ref{fig:spec}b corresponding 
closest to the observation position in the ecliptic plane. 
Only protons with velocities away from the Moon are included, 
to approximate the SWIM view conditions. 
The observed spreading of the solar wind spectrum to lower and 
higher energies is seen in the model spectrum. 
That the peak of the model spectrum is so much lower in energy 
than the solar wind (not seen in the observed spectrum), 
might indicate that the model reduction in velocity ($f_v=0.5$) 
is too large.  This is consistent with the SELENE observation that 
$f_v=0.8$~\citep{KaguyaGRL08}. 

\begin{figure}
 % mb22b5.   1.0 vel cos2spec (reflect_type = 6)
\begin{center}
  \includegraphics[width=1.0\columnwidth]{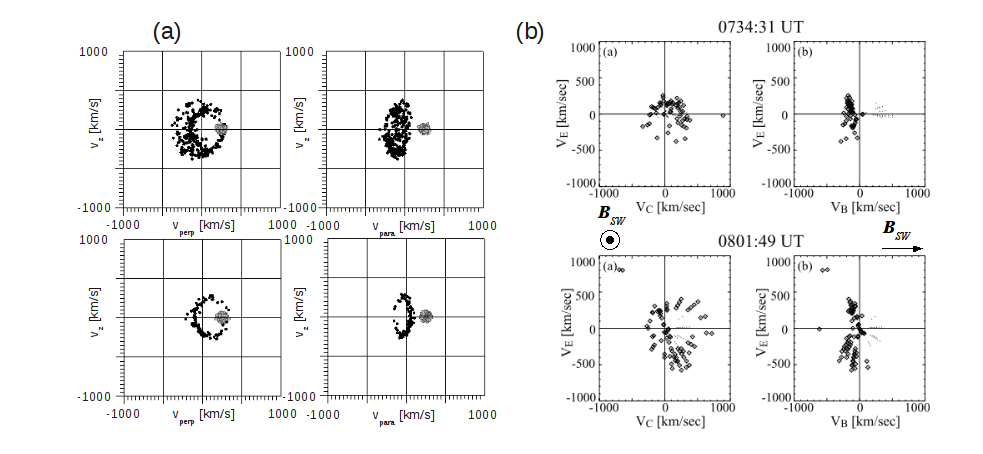}
\end{center}
\caption{The reflected protons in velocity space. 
Comparison of (a) test particle model at time 120~s, with 
(b) Nozomi observations. 
Top row is the velocity space distribution projected along $\mathbf{B}_{sw}$ 
(left) and perpendicular to $\mathbf{B}_{sw}$ (right). 
Shown are all simulation meta-particles in a cube with side of 500~km, 
centered at the positions of the observations, 
(2500,-3700,-900)~km and (-600,-5500,-1400)~km, respectively. 
These positions are shown in Fig.~\ref{fig:hybrid}. 
In gray are the solar wind protons, and in black are the reflected protons. 
The plots in (b) are from~\citet[Fig.~5]{Fut03}.  
The solar wind conditions used in the simulation has a velocity of
350~km/s, a temperature of 48000~K, and a magnetic field that is 
$(-2.82,1.03,0)$~nT. 
The reflection model is cos$^2$ specular. 
        } \label{fig:nozomi}
\end{figure}

We now use the test particle model to also investigate if reflected 
protons can explain the observations of 
non-thermal ions by the Nozomi mission. 
In~\citet{Fut03}, Fig.~5, ion velocity spectra taken at two different times 
during the Nozomi Lunar flyby are presented. 
To compare the results of the test particle model with Nozomi's 
ion observation, we must first of all select appropriate solar 
wind conditions. 
The estimated solar wind conditions in~\citep{Fut03} is 350~km/s
with an IMF of magnitude 3~nT, $(-2.12,2.12,0)$~nT. 
However, there are uncertainties in this 
IMF estimate since the direction was obtained from the observed 
temperature anisotropy in the electron velocity spectrum, 
and~\citet{Fut03} estimate the uncertainty in IMF direction to 
20$^\circ$. 
Due to this uncertainty, and since the trajectories of reflected protons 
are sensitive to the IMF, we tried different IMFs and found a best fit for 
$(-2.82,1.03,0)$~nT, i.e.\ about 25$^\circ$away from the estimate 
by~\citet{Fut03}. 
In Fig.~\ref{fig:nozomi} we compare velocity space plots 
of the test particle model with Nozomi's 
ion observation (Fig.~5 in~\citet{Fut03}). 

We see clear similarities in the observed and modeled 
velocity space distributions.  This shows that reflected protons 
is a possible explanation for the observed nonthermal ions. 
The discrepancies in distributions could be due to a change 
in the IMF between the two observation times.  
For an illustration of the observation geometry, see Fig.~2 in 
\citep{Fut03}. 
This show that the Nozomi ion observations, even upstream of the moon, 
can be explained by reflected solar wind ions. 

\begin{figure}
\begin{center}
  \includegraphics[width=0.8\columnwidth]{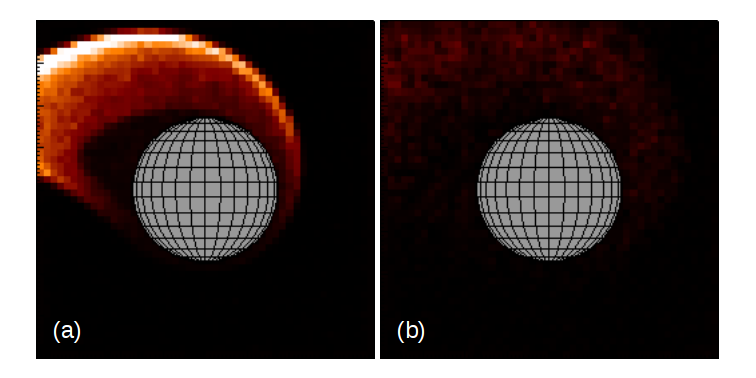}
  % (a) mb15 and (b) mb07
  %\vspace{10em}
\end{center}
\caption{Different reflection processes.  Here it is illustrated how 
the global solar wind interaction is perturbed by changes in the local 
microphysics of the reflection process.  Shown is the 
magnitude of the $yz$-component of the proton number flux, relative 
to the magnitude of the solar wind proton flux, in the plane $x=0$, when
the reflected protons are 
(a) perpendicular to the lunar surface with $f_v=0.5$, and 
(b) reflected according to the cos$^2$-specular model, 
    without losing velocity ($f_v=1$). 
The different reflection processes are illustrated in Fig.~\ref{fig:ill}d 
and the color scale is the same as in Fig.~\ref{fig:flux}a. 
All other parameters for these two simulation runs are the same as 
for that in Fig.~\ref{fig:flux}. 
        } \label{fig:proc}
\end{figure}

To illustrate the global effects of different assumptions for the 
local reflection process, Fig.~\ref{fig:proc} show the effect 
of perpendicular reflection, and of no velocity loss at reflection, $f_v=1$. 
The perpendicular reflection model give less velocity dispersion of 
the reflected ions, and lead to more than three times the solar wind 
number flux (assuming all protons are reflected).  
No velocity loss at reflection on the other hand lead to a larger 
velocity dispersion, and a more diffuse plume of ions, with only 
0.7 times the solar wind flux. 

\begin{figure}
% Two other good fits, mb22b and mb22b4 (reflect_type = 5 and 4). 
\begin{center}
  \includegraphics[width=1.0\columnwidth]{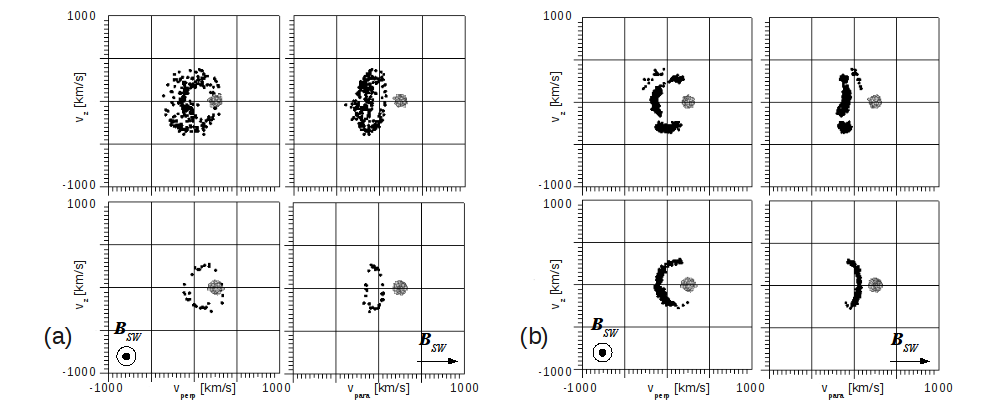}
\end{center}
\caption{The reflected protons in velocity space from the 
test particle model for the same Nozomi comparison as shown in 
Fig.~\ref{fig:nozomi}. 
This is a comparison of two different reflection models. 
(a) cos$^2$ perpendicular, and (b) specular. 
Both cases has the same solar wind conditions as in Fig.~\ref{fig:nozomi}. 
        } \label{fig:scatter} %\label{lastfig}
\end{figure}

To illustrate the sensitivity of the reflected protons to the 
microphysics of the reflection process, we show in Fig.~\ref{fig:scatter} 
how the velocity space distributions of the reflected protons change 
when we assume a cos$^2$ perpendicular and a specular reflection 
model. 

\section{Discussion}
We can compare the surface reflected ions with pick-up ions, 
observed at comets, Mars and Venus.  In a frame following the 
reflected ions, the situation is similar.  So a reflected 
proton behaves as a pick-up ions seeing a faster solar wind 
(from different directions for all reflected ions). 
This will create a ring distribution in velocity space, as 
was observed at the Moon~\citep{Fut03}. 
The cycloid trajectory of these reflected ions can take them 
deep into the wake of the Moon, 
which has been observed by SELENE~\citep{Nishino09b}, 
giving a possible explanation for 
the Apollo observations of solar wind energy 
ions at the night side surface of the Moon~\citep{fre72}. 
These were bursts of ions in the keV range, with fluxes of the order 
10$^6$~cm$^{-2}$ s$^{-1}$ sr$^{-1}$, and mass per unit charge less than 
10 amu/q, and hence thought to be of solar wind origin. 

It is interesting to note that a local process (surface reflection) 
can perturb the global interaction of the Moon with the solar 
wind.  Also, the character of this global interaction 
depends on the details of the local process, e.g., the 
velocity distribution of the reflected protons. 
Thus, it is possible to infer properties of the local 
reflection process from far away observations of the 
ion distributions. 

If we consider the energies involved in the reflection process, 
the kinetic energy density of the reflected ions is 
$W_r = f_r {n}_{sw} m_p (f_v{v}_{sw})^2/2$ 
where $m_p$ is the proton mass, % amu = 1.66054e-27 [kg]
and the magnetic field energy density is 
$W_B = {B}_{sw}^2/(2\mu_0)$
where $\mu_0$ % $\mu_0 = 4\pi\cdot 10^{-7}$~N/A$^2$
is the magnetic constant. 
Using $f_r=0.01$, $n_{sw}=5$~cm$^{-3}$, $f_v=0.5$, 
$v_{sw}=350$~km/s, and $B_{sw}$ = 3~nT, 
we get that $W_r/W_B$ = 0.35. 
%; 0.01*5e6*1.66e-27*(0.5*350e3)^2/2
%	0.0000000000012709375
%; (3e-9)^2/(2*4*3.14e-7)
%	~0.00000000000358280255
%; 0.0000000000012709375/0.00000000000358280255
%	~0.35473277755705516063
Thus, the kinetic energy density of the reflected ions is a 
significant fraction of the solar wind magnetic field energy density, 
and the reflected ions should be a strong source of 
wave activity. 

We can compare these reflected protons with the protons reflected 
by Earth's bow shock.  There, in Earth's foreshock region, these 
ion beams propagate upstream in the solar wind.  
It is a classical case of an electromagnetic counter-streaming beam 
situation, with the beam density around 1\% of the solar wind density. 
This causes wave activity, and has been studied for a long time in 
great detail, see e.g.~\citet{tsu81}. 

The reflection of solar wind ions should be a universal process 
that occur at all bodies without a significant atmosphere, e.g., 
at asteroids, and at the Martian moons Phobos and Deimos. 

\section*{Acknowledgments}
This research was conducted using the resources of the 
High Performance Computing Center North (HPC2N), Ume\aa\ University, Sweden, 
and the Center for Scientific and Technical Computing (LUNARC), 
Lund University, Sweden. 
The software used in this work was in part developed by the 
DOE-supported ASC / Alliance Center for Astrophysical 
Thermonuclear Flashes at the University of Chicago.

\bibliography{hybrid}

\begin{thebibliography}{25}
\providecommand{\natexlab}[1]{#1}
\providecommand{\url}[1]{\texttt{#1}}
\expandafter\ifx\csname urlstyle\endcsname\relax
  \providecommand{\doi}[1]{doi: #1}\else
  \providecommand{\doi}{doi: \begingroup \urlstyle{rm}\Url}\fi

\bibitem[Barabash et~al.(2009)]{bar09}
Stas Barabash et~al.
\newblock Investigation of the solar wind--{M}oon interaction onboard
  {C}handrayaan-1 mission with the {SARA} experiment.
\newblock \emph{Current Science}, 96\penalty0 (4):\penalty0 526--532, 2009.

\bibitem[{Bhardwaj} et~al.(2005){Bhardwaj}, {Barabash}, {Futaana}, {Kazama},
  {Asamura}, {McCann}, {Sridharan}, {Holmstrom}, {Wurz}, and {Lundin}]{bha05}
A.~{Bhardwaj}, S.~{Barabash}, Y.~{Futaana}, Y.~{Kazama}, K.~{Asamura},
  D.~{McCann}, R.~{Sridharan}, M.~{Holmstrom}, P.~{Wurz}, and R.~{Lundin}.
\newblock {Low energy neutral atom imaging on the Moon with the SARA instrument
  aboard Chandrayaan-1 mission}.
\newblock \emph{Journal of Earth System Science}, 114:\penalty0 749--760,
  December 2005.
\newblock \doi{10.1007/BF02715960}.

\bibitem[Cravens(2004)]{Cravens}
Thomas~E. Cravens.
\newblock \emph{Physics of Solar System Plasmas}.
\newblock Cambridge University Press, 2004.

\bibitem[Crider and Vondrak(2002)]{cri02}
D.~H. Crider and R.~R. Vondrak.
\newblock Hydrogen migration to the lunar poles by solar wind bombardment of
  the moon.
\newblock \emph{Advances in Space Research}, 30\penalty0 (8):\penalty0
  1869--1874, 2002.

\bibitem[{Farrell} et~al.(1998){Farrell}, {Kaiser}, {Steinberg}, and
  {Bale}]{Farrell98}
W.~M. {Farrell}, M.~L. {Kaiser}, J.~T. {Steinberg}, and S.~D. {Bale}.
\newblock {A simple simulation of a plasma void: Applications to Wind
  observations of the lunar wake}.
\newblock \emph{Journal of Geophysical Research}, 103:\penalty0 23653--23660,
  1998.
\newblock \doi{10.1029/97JA03717}.

\bibitem[{Freeman}(1972)]{fre72}
J.~W. {Freeman}, Jr.
\newblock {Energetic ion bursts on the nightside of the moon.}
\newblock \emph{Journal of Geophysical Research}, 77:\penalty0 239--243, 1972.
\newblock \doi{10.1029/JA077i001p00239}.

\bibitem[{Futaana} et~al.(2003){Futaana}, {Machida}, {Saito}, {Matsuoka}, and
  {Hayakawa}]{Fut03}
Y.~{Futaana}, S.~{Machida}, Y.~{Saito}, A.~{Matsuoka}, and H.~{Hayakawa}.
\newblock {Moon-related nonthermal ions observed by Nozomi: Species, sources,
  and generation mechanisms}.
\newblock \emph{Journal of Geophysical Research (Space Physics)}, 108:\penalty0
  1025, 2003.
\newblock \doi{10.1029/2002JA009366}.

\bibitem[Goswami and Annadurai(2009)]{gos09}
J.N.\ Goswami and M.\ Annadurai.
\newblock {C}handrayaan-1: {I}ndia's first planetary science mission to the
  moon.
\newblock \emph{Current Science}, 96\penalty0 (4):\penalty0 486--491, 2009.

\bibitem[{Halekas} et~al.(2005){Halekas}, {Bale}, {Mitchell}, and
  {Lin}]{Halekas05}
J.~S. {Halekas}, S.~D. {Bale}, D.~L. {Mitchell}, and R.~P. {Lin}.
\newblock {Electrons and magnetic fields in the lunar plasma wake}.
\newblock \emph{Journal of Geophysical Research (Space Physics)}, 110\penalty0
  (A9):\penalty0 7222, 2005.
\newblock \doi{10.1029/2004JA010991}.

\bibitem[{Harnett} and {Winglee}(2003)]{har03}
E.~M. {Harnett} and R.~M. {Winglee}.
\newblock {2.5-D fluid simulations of the solar wind interacting with multiple
  dipoles on the surface of the Moon}.
\newblock \emph{Journal of Geophysical Research (Space Physics)}, 108:\penalty0
  1088, 2003.
\newblock \doi{10.1029/2002JA009617}.

\bibitem[Holmstr\"{o}m(2009)]{Enumath09}
M.~Holmstr\"{o}m.
\newblock Hybrid modeling of plasmas.
\newblock To appear in the Proceedings of the 8th European Conference:
  Numerical Mathematics and Advanced Applications, Springer.
  http://arxiv.org/abs/0911.4435, 2009.

\bibitem[Kallio(2005)]{kal05}
E.\ Kallio.
\newblock Formation of the lunar wake in quasi-neutral hybrid model.
\newblock \emph{Geophysical Research Letters}, 32:\penalty0 L06107, 2005.

\bibitem[Kimura and Nakagawa(2008)]{kimura08}
Shinya Kimura and Tomoko Nakagawa.
\newblock Electromagnetic full particle simulation of the electric field
  structure around the moon and the lunar wake.
\newblock \emph{Earth Planets Space}, 60:\penalty0 591--599, 2008.

\bibitem[{Manka} and {Michel}(1970)]{Manka70}
R.~H. {Manka} and F.~C. {Michel}.
\newblock {Lunar Atmosphere as a Source of Argon-40 and Other Lunar Surface
  Elements}.
\newblock \emph{Science}, 169:\penalty0 278--280, July 1970.

\bibitem[{Manka} and {Michel}(1973)]{Manka73}
R.~H. {Manka} and F.~C. {Michel}.
\newblock {Lunar ion energy spectra and surface potential}.
\newblock In \emph{Lunar and Planetary Science Conference}, volume~4 of
  \emph{Lunar and Planetary Science Conference}, pages 2897--2908, 1973.

\bibitem[{McCann} et~al.(2007){McCann}, {Barabash}, {Nilsson}, and
  {Bhardwaj}]{swim}
D.~{McCann}, S.~{Barabash}, H.~{Nilsson}, and A.~{Bhardwaj}.
\newblock {Miniature ion mass analyzer}.
\newblock \emph{Planetary and Space Science}, 55:\penalty0 1190--1196, June
  2007.
\newblock \doi{10.1016/j.pss.2006.11.020}.

\bibitem[{Nishino} et~al.(2009{\natexlab{a}}){Nishino}, {Fujimoto}, {Maezawa},
  {Saito}, {Yokota}, {Asamura}, {Tanaka}, {Tsunakawa}, {Matsushima},
  {Takahashi}, {Terasawa}, {Shibuya}, and {Shimizu}]{Nishino09b}
M.~N. {Nishino}, M.~{Fujimoto}, K.~{Maezawa}, Y.~{Saito}, S.~{Yokota},
  K.~{Asamura}, T.~{Tanaka}, H.~{Tsunakawa}, M.~{Matsushima}, F.~{Takahashi},
  T.~{Terasawa}, H.~{Shibuya}, and H.~{Shimizu}.
\newblock {Solar-wind proton access deep into the near-Moon wake}.
\newblock \emph{Geophysical Research Letters}, 36:\penalty0 16103, August
  2009{\natexlab{a}}.
\newblock \doi{10.1029/2009GL039444}.

\bibitem[{Nishino} et~al.(2009{\natexlab{b}}){Nishino}, {Maezawa}, {Fujimoto},
  {Saito}, {Yokota}, {Asamura}, {Tanaka}, {Tsunakawa}, {Matsushima},
  {Takahashi}, {Terasawa}, {Shibuya}, and {Shimizu}]{Nishino09a}
M.~N. {Nishino}, K.~{Maezawa}, M.~{Fujimoto}, Y.~{Saito}, S.~{Yokota},
  K.~{Asamura}, T.~{Tanaka}, H.~{Tsunakawa}, M.~{Matsushima}, F.~{Takahashi},
  T.~{Terasawa}, H.~{Shibuya}, and H.~{Shimizu}.
\newblock {Pairwise energy gain-loss feature of solar wind protons in the
  near-Moon wake}.
\newblock \emph{Geophysical Research Letters}, 36:\penalty0 12108, June
  2009{\natexlab{b}}.
\newblock \doi{10.1029/2009GL039049}.

\bibitem[{Ogilvie} et~al.(1996){Ogilvie}, {Steinberg}, {Fitzenreiter}, {Owen},
  {Lazarus}, {Farrell}, and {Torbert}]{Ogilvie96}
K.~W. {Ogilvie}, J.~T. {Steinberg}, R.~J. {Fitzenreiter}, C.~J. {Owen}, A.~J.
  {Lazarus}, W.~M. {Farrell}, and R.~B. {Torbert}.
\newblock {Observations of the lunar plasma wake from the WIND spacecraft on
  December 27, 1994}.
\newblock \emph{Geophysical Research Letters}, 23:\penalty0 1255--1258, 1996.
\newblock \doi{10.1029/96GL01069}.

\bibitem[Saito et~al.(2008)Saito, Yokota, Tanaka, Asamura, Nishino, Fujimoto,
  Tsunakawa, Shibuya, Matsushima, Shimizu, Takahashi, Mukai, and
  Terasawa]{KaguyaGRL08}
Y.~Saito, S.~Yokota, T.~Tanaka, K.~Asamura, M.~N. Nishino, M.~Fujimoto,
  H.~Tsunakawa, H.~Shibuya, M.~Matsushima, H.~Shimizu, F.~Takahashi, T.~Mukai,
  and T.~Terasawa.
\newblock {Solar wind proton reflection at the lunar surface: Low energy ion
  measurement by MAP-PACE onboard SELENE (KAGUYA)}.
\newblock \emph{Geophysical Research Letters}, 35:\penalty0 L24205, 2008.

\bibitem[{Schubert} and {Lichtenstein}(1974)]{shu74}
G.~{Schubert} and B.~R. {Lichtenstein}.
\newblock {Observations of moon-plasma interactions by orbital and surface
  experiments}.
\newblock \emph{Reviews of Geophysics and Space Physics}, 12:\penalty0
  592--626, November 1974.

\bibitem[{Shimazu}(1999)]{Shimazu99}
H.~{Shimazu}.
\newblock {Three-dimensional hybrid simulation of magnetized plasma flow around
  an obstacle}.
\newblock \emph{Earth, Planets, and Space}, 51:\penalty0 383--393, 1999.

\bibitem[{Tr{\'a}vn{\'{\i}}{\v c}ek} et~al.(2005){Tr{\'a}vn{\'{\i}}{\v c}ek},
  {Hellinger}, {Schriver}, and {Bale}]{tra05}
P.~{Tr{\'a}vn{\'{\i}}{\v c}ek}, P.~{Hellinger}, D.~{Schriver}, and S.~D.
  {Bale}.
\newblock {Structure of the lunar wake: Two-dimensional global hybrid
  simulations}.
\newblock \emph{Geophysical Research Letters}, 32:\penalty0 L06102, 2005.

\bibitem[{Tsurutani} and {Rodriguez}(1981)]{tsu81}
B.~T. {Tsurutani} and P.~{Rodriguez}.
\newblock {Upstream waves and particles - An overview of ISEE results}.
\newblock \emph{Journal of Geophysical Research}, 86:\penalty0 4319--4324,
  1981.
\newblock \doi{10.1029/JA086iA06p04317}.

\bibitem[Wieser et~al.(2009)Wieser, Barabash, Futaana, Holmstr\"{o}m, Bhardwaj,
  Sridharan, Dhanya, Wurz, Schaufelberger, and Asamura]{wie09}
Martin Wieser, Stas Barabash, Yoshifumi Futaana, Mats Holmstr\"{o}m, Anil
  Bhardwaj, R.~Sridharan, M.B. Dhanya, Peter Wurz, Audrey Schaufelberger, and
  Kazushi Asamura.
\newblock Extremely high reflection of solar wind protons as neutral hydrogen
  atoms from regolith in space.
\newblock \emph{Planetary and Space Science}, 57\penalty0 (14-15):\penalty0
  2132--2134, 2009.

\end{thebibliography}
\bibliographystyle{plainnat} 

\label{lastpage}

\end{document}